\documentclass[aps,pra,twocolumn]{revtex4}
\usepackage{amssymb}

\usepackage[pctex32]{graphics}

\begin{document}

\title{Adiabatic and non-adiabatic merging of independent Bose condensates}
\author{W. Yi, and L.-M. Duan \\
\textit{\small FOCUS Center and MCTP, Department of Physics, University of
Michigan, Ann Arbor, MI 48109}}

\begin{abstract}
Motivated by a recent experiment [Chikkatur \textit{et al.} \textit{Science}%
, \textbf{296}, 2193 (2002)] on the merging of atomic condensates, we
investigate how two independent condensates with random initial phases can
develop a unique relative phase when we move them together. In the adiabatic
limit, the uniting of independent condensates can be understood from the
eigenstate evolution of the governing Hamiltonian, which maps degenerate
states (corresponding to fragmented condensates) to a single state
(corresponding to a united condensate) . In the non-adiabatic limit
corresponding to the practical experimental configurations, we give an
explanation on why we can still get a large condensate fraction with a
unique relative phase. Detailed numerical simulations are then performed for
the non-adiabatic merging of the condensates, which confirm our explanation
and qualitative estimation. The results may have interesting implications
for realizing a continuous atom laser based on merging of condensates.
\end{abstract}

\maketitle

\address{FOCUS center and MCTP, Department of Physics, University of Michigan, Ann
Arbor, MI 48109}

\section{Introduction}

The prospect of the creation of a continuous, coherent atomic beam, the
atomic analogy of laser beam, has been of great interest to many ever since
the successful generation of Bose-Einstein condensate (BEC) \cite
{1,2,16,3,13,14,15,17}. One of the major challenges to such applications
lies in the difficulty in continuous condensation of atomic gas due to
stringent cooling conditions. Towards that goal, one needs to spatially
separate the evaporative cooling from the destructive laser cooling, which
is still pretty challenging even though there've been interesting
investigations and progresses \cite{17,18,19}. Alternatively, one can
realize a continuous source of condensate by bringing new condensates into
the trap and uniting them. Note that independently prepared condensates have
a completely random relative phase; in order to unite them, one needs to
have a phase cooling mechanism to get rid of this random phase. Interesting
theoretical investigations have been carried out on the possibilities of
uniting two existing condensates through laser induced phase cooling with
condensates confined in a high-Q ring cavity \cite{4}, or through phase
locking with feedbacks from a series of interference measurements \cite{5}.

Recently, Chikkatur \textit{et al.} reported a striking experiment in which
two independently produced condensates were merged directly in space by
bringing their traps together to a complete overlap\cite{6}. A condensate
fraction larger than the initial components was observed from the subsequent
time-of-flight imaging. This raises the question as how the relative phase
between the two component condensates is established during this direct
merging where there is no additional phase cooling or locking. In this
paper, we investigate this puzzling phase dynamics both in the ideal
adiabatic merging limit and in the more practical non-adiabatic
circumstance. In the ideal adiabatic limit, we show that a unique relative
phase will be established between the two components during direct merging,
resulting from the eigenstate evolution of the governing Hamiltonian (an
effective two-mode model). The Hamiltonian has degenerate ground states
corresponding to two fragmented condensates when the traps are apart. When
we overlap the two traps, the degenerate ground states evolve into a single
eigenstate corresponding to a single condensate with a unique relative phase
between the two initial components. While the adiabatic limit requires very
slow merging, the real experiment is actually performed far from that limit
\cite{6}. In this case, the atoms will slip from the ground state to all the
nearby eigenstates, and one would not expect to get a single condensate.
However, through a careful analysis of the eigenstate structure of the
governing Hamiltonian, we argue that one can still get a large condensate
fraction (larger than either of the initial components') with a unique
relative phase in this non-adiabatic limit. We then confirm this evolution
picture through detailed numerical simulations of the merging dynamics.

We should also point out that the non-adiabatic merging considered in this
paper, which is based on the parameters of the experiment \cite{6}, is not
an abrupt connection of the two condensates. In the latter case, a unique
relative phase cannot be established without careful consideration of
dissipation \cite{7}. In the experiment, the time scale of the merger(~0.5
s) is long when compared with the trap frequency (440Hz) along the merging
direction\cite{6}. So the merging is actually adiabatic with respect to the
evolution of the condensate wave packets; and dissipation due to the
quasi-particle excitation in each well should not play an important role.
However, this same merging time scale is very short when compared with the
lowest excitation energy of the relevant Hamiltonian (the effective two-mode
model). Many eigenstates of this Hamiltonian will get populated during the
merger. That is why we call it "non-adiabatic". Our central observation is
that due to the particular eigen-energy distribution of this relevant
two-mode Hamiltonian, even for mergers much faster than the lowest
excitation energy, only low-lying states of the Hamiltonian will be
populated. Furthermore, for all the populated low-lying states, the relative
phase between the two initial condensate components is well locked. As a
result, we will get a large merged condensate fraction with a uniquely
established relative phase.

In the following, we first present a theoretical model for the description
of the direct merging process, and argue that the internal phase dynamics is
mainly captured by the simplified two-mode Hamiltonian with time dependent
parameters, which has been a popular model for many theoretical
investigations in different scenarios \cite{7,8,9,10,11,12}. After that, we
go to the main topic and investigate the relative phase dynamics. The result
in the adiabatic limit can be easily understood. In the non-adiabatic limit,
the arguments need to be based on the analysis of the detailed evolution
structure of the eigenstates of the time-dependent Hamiltonian. Then in Sec.
III, we investigate the relative phase dynamics and the evolution of the
condensate fraction through detailed numerical simulations, with the
evolution speed ranging from the adiabatic limit to the far non-adiabatic
circumstance. We simulate the merging process with different configurations
and different initial states of the two components. Finally, we discuss the
relevance of the results to the reported experiment and to the prospect of
producing continuous atom lasers based on condensate merging.

\section{Theoretical modeling of adiabatic and non-adiabatic merging of
independent condensates}

\subsection{An effective two-mode model for the description of the relative
phase dynamics}

Consider two independently prepared condensates in separate traps at low
temperature. The two traps are then brought closer along a certain direction
to allow the condensates to merge, as illustrated in Fig. 1. The traps can
be modeled as a time-dependent double-well potential $V_{T}(\mathbf{r},t)$,
with the two wells moving towards each other. The interactions between the
atoms are described by the usual $\delta $-pseudopotential, $U(\mathbf{%
r_{1},r_{2}})=(4\pi \hbar a_{s}/m)\delta (\mathbf{r_{1},r_{2}})$, where $%
a_{s}$ is the s-wave scattering length and $m$ is the mass of the atom. The
second quantization Hamiltonian for such a system has the form
\begin{eqnarray}
\hat{H} &=&\int \hat{\Psi}^{\dag }(\mathbf{r})(\frac{P^{2}}{2m}+V_{T}(%
\mathbf{r},t))\hat{\Psi}(\mathbf{r})d\mathbf{r}  \nonumber \\
&&+\frac{g}{2}\int \hat{\Psi}^{\dag }(\mathbf{r})\hat{\Psi}^{\dag }(\mathbf{r%
})\hat{\Psi}(\mathbf{r})\hat{\Psi}(\mathbf{r})d\mathbf{r},  \label{1}
\end{eqnarray}
where $\hat{\Psi}\left( r\right) $ is the bosonic quantum field operator
describing the atoms, and the parameter $g=4\pi \hbar a_{s}/m$.
\begin{figure}[h]
\includegraphics{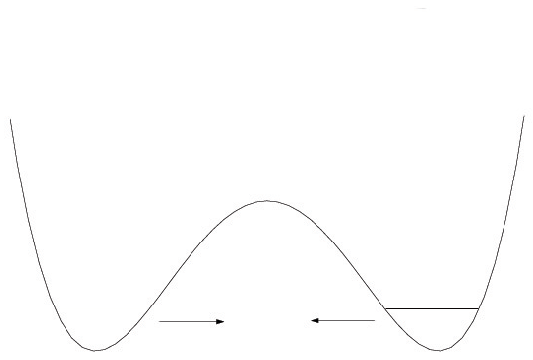}
\caption[Fig.1 ]{Illustration of merging of two independent condensates.}
\end{figure}

When the traps are not overlapping, the atoms in the same condensate are in
the same quantum mode. We can expand the atomic field operator $\hat{\Psi}(%
\mathbf{r})$ as $\hat{\Psi}(\mathbf{r})=a_{1}\phi _{1}(\mathbf{r})+a_{2}\phi
_{2}(\mathbf{r}),$, where $\phi _{i}(\mathbf{r})$ $\left( i=1,2\right) $\
are the corresponding mean-fields from the Gross-Pitaevski equation for each
trap. When we bring the two condensates closer by varying the trap potential
$V_{T}(\mathbf{r},t)$, we assume that the atomic density profile
adiabatically follows the movement of the traps, as is the case in the
experiment\cite{6}. With this assumption, at any given time $t$, we can
still use a variational mean field $\phi _{i}(\mathbf{r,t})$ to describe the
corresponding condensate. The condensate wave functions now overlap in
space, and the atoms in different condensates can tunnel to each other
through the trap barrier. As a good approximation, we can expand the atomic
field $\hat{\Psi}(\mathbf{r,t})$ at time $t$ as $\hat{\Psi}(\mathbf{r,t}%
)=a_{1}\phi _{1}(\mathbf{r,t})+a_{2}\phi _{2}(\mathbf{r,t})$. With this
expansion, the Hamiltonian (1) is simplified to the following well-known
two-mode model
\begin{equation}
\hat{H}=\frac{U\left( t\right) }{4}\left( a_{2}^{\dag }a_{2}-a_{1}^{\dag
}a_{1}\right) ^{2}-\frac{J\left( t\right) }{2}(a_{1}^{\dag
}a_{2}+a_{2}^{\dag }a_{1}),  \label{2}
\end{equation}
where the parameters $U\left( t\right) $ and $J\left( t\right) $ are given
by
\begin{equation}
U(t)=\frac{g}{2}\int (\left| \phi _{1}\right| ^{4}+\left| \phi _{2}\right|
^{4}-2\left| \phi _{1}\right| ^{2}\left| \phi _{2}\right| ^{2})d\mathbf{r},
\label{6}
\end{equation}
\begin{equation}
J(t)=-2\int \phi _{1}^{\ast }(\frac{P^{2}}{2M}+V(\mathbf{r},t))\phi _{2}d%
\mathbf{r}.  \label{7}
\end{equation}
In writing Eq. (2), we have assumed a symmetric trap with $\int \left| \phi
_{1}\right| ^{4}d\mathbf{r}=\int \left| \phi _{2}\right| ^{4}d\mathbf{r}$,
and we have neglected terms that are functions only of the atom number $\hat{%
N}$, as they commute with $\hat{H}$ and can be eliminated by going
to a rotating frame. We have also neglected the nonlinear
tunneling terms proportional to $(a_{1}^{\dag }a_{2}+a_{2}^{\dag
}a_{1})^2$ (see Ref. \cite {10} for the coefficient before that
term) because they are either typically smaller than or have the
same effect as the linear tunneling, so that neglecting them does
not change the basic physics \cite {21}. Note that given all the
approximations above, the factor before the last term in Eq. (3)
has some arbitrariness as this term is on the same order of
magnitude as some neglected terms in the weak overlapping limit.
As a convenient choice, we take the factor to be $2$ so that we
have a vanishing $U(t)$ as the two traps completely overlap, as
one would expect on physical grounds. The two-mode Hamiltonian
(2), which is a very popular model for many theoretical
investigations in different scenarios \cite{7,8,9,10,11,12},
catches the main physics of condensates in double-well potentials
resulting from the competition between the tunneling and the local
nonlinear collision interaction. A more detailed and more rigorous
derivation of the two-mode model from the variational
approximation can be found in Ref. \cite{12}. One important
limitation for applications of the two-mode model is that when the
two condensates get very close, the coupling between the two
condensates becomes so strong that some approximations underlying
the two-mode model are not well justified. In that strong
overlapping region, Eqs. (2)-(4) serve as a convenient
extrapolation. However, numerical calculation reveals that the
relative phase and the condensate fraction are established shortly
after the commencement of the merger, when the two condensates are
not yet strongly coupled. It is under this context that we think
an effective two-mode model catches the main physics and is
adequate for the understanding of the relative phase evolution.

We will use the time-dependent two-mode model (Hamiltonian (2)) to
investigate the relative phase dynamics during the merging of independent
condensates. The temporal behavior of the coefficients $U\left( t\right) $
and $J\left( t\right) $ in principle should be derived from the
self-consistent variational solutions of $\phi _{1}(\mathbf{r,t})$ and $\phi
_{2}(\mathbf{r,t})$, which is pretty involved. Fortunately, in the following
we will see that the relative phase dynamics is actually insensitive to the
details of $U\left( t\right) $ and $J\left( t\right) $. It only depends on
the rough scale of their rates of variation. Hence, for the following
discussions in this section, we do not specify the explicit forms of $%
U\left( t\right) $ and $J\left( t\right) $. Nevertheless, we do know the
rough trends of the evolutions of $U\left( t\right) $ and $J\left( t\right) $%
. As the two traps approach each other, $J\left( t\right) $\ will
continuously increase from zero to a maximum value $J_{0}$. Compared with $%
J(t)$, $U\left( t\right) $ typically changes much more slowly at the
beginning of the merger, and can be roughly taken as a constant $U_{0}.$ As
the two traps further merge approaching to a complete overlap, one expects
that $U\left( t\right) $ will gradually decrease to zero. These rough trends
of evolutions of $U\left( t\right) $ and $J\left( t\right) $ are important
for the understanding of the relative phase dynamics.

\subsection{Merging of condensates in the adiabatic limit}

For the convenience of the following discussions, we write the two-mode
model (2) using effective collective spin operators. With the standard
Schwinger representation of the spin operators $\hat{S}_{x}=\frac{1}{2}%
(a_{1}^{\dag }a_{2}+a_{2}^{\dag }a_{1})$, $\hat{S}_{y}=\frac{i}{2}%
(a_{1}^{\dag }a_{2}-a_{2}^{\dag }a_{1})$, and $\hat{S}_{z}=\frac{1}{2}%
(a_{2}^{\dag }a_{2}-a_{1}^{\dag }a_{1})$, Hamiltonian (2) can be expressed
in the following form
\begin{equation}
\hat{H}\left( t\right) =U\left( t\right) \hat{S}_{z}^{2}-J\left( t\right)
\hat{S}_{x}.
\end{equation}
This Hamiltonian is actually equivalent to that of the infinite-range Ising
model \cite{20}, and one has effective anti-ferromagnetic interaction as $%
U\left( t\right) >0$.

First, let us assume that the temperature of the atoms is low enough, and
that the variation speeds of $U\left( t\right) $ and $J\left( t\right) $ are
very small. In this ideal adiabatic limit, the establishment of a unique
relative phase through direct merging of condensates can be easily
understood from the adiabatic theorem, which states that the system in such
a limit will remain in the ground state of the Hamiltonian $\hat{H}\left(
t\right) $. Initially, the two traps are far apart, therefore the self
interaction among the atoms in each trap dominates with $U\left( t\right)
\rightarrow U_{0}$ and $J\left( t\right) \rightarrow 0 $. The ground state
in this case is an eigenstate of $\hat{S}_{z}^{2}$ with eigenvalue of $0$,
i.e., with $\left\langle a_{2}^{\dag }a_{2}\right\rangle=\left\langle
a_{1}^{\dag }a_{1}\right\rangle=N/2$, where $N$ is the total atom number.
The two condensates have equal number of atoms, but there is no phase
relation between them, which corresponds to fragmented condensates. In the
final stage of the merger, the coefficients $U\left( t\right) \rightarrow 0$
and $J\left( t\right) \rightarrow J_{0}$, the coupling energy $-J\left(
t\right) \hat{ S_{x}}$ thus dominates in the Hamiltonian. The ground state
in this case is an eigenstate of $\hat{S}_{x}$ with the largest eigenvalue $%
N/2$ (without loss of generality we assume $J\left( t\right)>0$). This state
is actually an eigenstate of the number operator $a_{+}^{\dagger }a_{+}$ of
the mode $a_{+}=\left( a_{1}+a_{2}\right) /\sqrt{2}$ with an eigenvalue of $N
$, which corresponds to a single condensate. We then have a unique zero
relative phase between the two initial condensate modes $a_{1}$ and $a_{2}$,
which is due to the fact that the coupling energy $-J\left( t\right) \hat{%
S_{x}}$ is minimized only when we have the same phase between $a_{1} $ and $%
a_{2}$.

The above understanding of the relative phase establishment from the
adiabatic theorem seems to be simple and straightforward, but in a sense it
is pretty nontrivial. The relative phase really comes from a joint effect of
the two competitive terms in the Hamiltonian (5). If we have only the
collision interaction term $U\left( t\right) \hat{S}_{z}^{2}$ and start with
fragmented condensates with $\left\langle a_{2}^{\dag
}a_{2}\right\rangle=\left\langle a_{1}^{\dag }a_{1}\right\rangle=N/2$, it's
obvious that the condensate fractions will never be changed by this
collision interaction. On the other hand, if we have only the coupling term $%
-J\left( t\right) \hat{S_{x}}$, it will only rotate the two condensate modes
$a_{1}$ and $a_{2}$ to two different modes $a_{1}^{\prime }$ and $%
a_{2}^{\prime }$, and will never change the condensate fractions either. The
condensate fractions can be quantitatively described by the eigenvalues of
the following single-particle density matrix \cite{12}
\begin{equation}
\rho =\left(
\begin{array}{cc}
\left\langle a_{1}^{\dag }a_{1}\right\rangle & \left\langle a_{1}^{\dag
}a_{2}\right\rangle \\
\left\langle a_{2}^{\dag }a_{1}\right\rangle & \left\langle a_{2}^{\dag
}a_{2}\right\rangle
\end{array}
\right) .  \label{9}
\end{equation}
The system is in a single-condensate state if $\rho $ has only one
macroscopic eigenvalue in the limit $N\rightarrow \infty $; and it
corresponds to fragmented condensates if $\rho $ has more macroscopic
eigenvalues in the large $N$ limit. It is clear that the coupling term $%
-J\left( t\right) \hat{S_{x}}$ only makes a linear rotation of the modes $%
a_{1}$ and $a_{2}$, and will never change the eigenvalues of the
single-particle density matrix $\rho $. Clearly, it is the
interplay between these two competing interactions in the
Hamiltonian (5) that is really responsible for producing a larger
condensate fraction with a well-defined relative phase.

The analysis above gives us a simple understanding of how a unique
relative phase in principle could be established between two
initially independent condensates through direct merging in the
perfect adiabatic limit. However, this analysis hardly explains
the merging dynamics in real experiments because the practical
configurations are far from this ideal adiabatic limit.
Practically, first of all, we can never start with an equal number
state for the two condensates. One can at most control the number
of atoms in each condensate so that they are roughly the same, as
each condensate has some number fluctuations typically on the
order of $\sqrt{N}$. So in reality we can not start from the
ground state of the Hamiltonian (5), but rather a statistical
mixture of some low-lying states. Secondly, to fulfill the
conditions of the adiabatic passage, one requires that the
variation speeds of $U\left( t\right) $ and $J\left( t\right) $ be
significantly smaller than the smallest eigen-frequency splitting
of the relevant Hamiltonian. The Hamiltonian (5) has an energy
difference of $U_{0}$ between the ground and the first excited
state when $U\left( t\right)\gg J\left( t\right) $ (i.e., when the
traps are far apart). This is a tiny energy gap, which can be
roughly estimated by $\mu _{c}/N$, where $\mu _{c}$ denotes the
chemical potential of the condensate ($\sim 2$kHz \cite{21}) and
the atom number is typically around $10^{6}$. The merging time
scale in real experiments ($\sim 0.5$ s \cite{6}) is much shorter
than the time scale given by $1/U_{0}$. This shows that we need to
go beyond the adiabatic limit to understand the merging dynamics
in practical configurations, which is the focus of our next
subsection.

\subsection{Merging of condensates in the practical non-adiabatic
circumstance}

If the evolution speed of the Hamiltonian is beyond the adiabatic
limit, the system prepared in the ground state will be excited to
some upper eigenstates. The excitation probability will depend on
the ratio of the variation rate of the Hamiltonian to the smallest
energy difference between the two eigenstates \cite{23}. The final
state in this case is in general a superposition of different
low-lying eigenstates which have been populated during the merger.
Therefore, to find out the final state of the system in the
non-adiabatic circumstance, it is important to look at the
evolution of all the eigenstates of the Hamiltonian (5) during the
merging process.

We have calculated the energy spectrum of the Hamiltonian (5) as a
function of the ratio of $J\left( t\right) /U\left( t\right) $ for
various atom numbers up to a few hundred (see also Ref. \cite{24},
where the energy spectrum is drawn for other purposes). The
evolutions of the energy spectrum for different atom numbers are
qualitatively very similar. To have a clear display, we only plot
in Fig. 2 the whole energy spectrum of the Hamiltonian (5) with a
small atom number $N=20$, as the energy levels become too dense to
be seen clearly in a small figure if $N$ is large. Fig. 2 is
enough to show some general properties of the energy spectrum
which are critical for the understanding of the merging dynamics
outside of the adiabatic limit. The spectrum can be divided into
three regions, corresponding respectively to the Fock, the
Josephson, and the Rabi region, as discussed in general
Josephson physics \cite{21}. First of all, on the left side of the spectrum (%
$J\left( t\right) \rightarrow 0$), which corresponds to the Fock
region, there are a total of $N+1$ eigenlevels. While the lowest
eigenlevel ($j=0$)
stays non-degenerate, the $\left( 2j-1\right) $th and $\left( 2j\right) $th (%
$j=1,2,\cdots ,N/2$) levels are degenerate in energy. The two degenerate
states can be expressed as $\left| N/2+j,N/2-j\right\rangle $ and $\left|
N/2-j,N/2+j\right\rangle $ in the Fock basis of the modes $a_{1}$ and $a_{2}$
. The energy difference between the ground state and the $2j$th eigenlevel
is given by $j^{2}U\left( t\right) \approx j^{2}U_{0}$. The quadratic
scaling of the energy difference with $j$ is very important for the
following discussions. Secondly, on the far right side of the spectrum with $%
J\gg NU$ (not shown in Fig. 2), the system enters the Rabi region,
where the eigenlevels are almost equally spaced. In that region,
one can approximately neglect the term $U\left( t\right)
J_{z}^{2}$, and the energy levels are
roughly eigenstates of the $\hat{S_{x}}$ operator. The $j$th ($%
j=0,1,2,\cdots N$) eigenlevel, expressed as $\left|
S=N/2,S_{x}=N/2-j\right\rangle $ in the Dicke basis of the
collective spin operator $\overrightarrow{S}$, has the energy
$jJ\left( t\right) \approx jJ_{0}$ (compared with the ground
state). Between these two limiting cases, there is a wide
intermediate Josephson region where the two-mode model could be
understood semi-classically \cite{21}.  In the Josephson region,
with fixed $J$ and $U$, the energy levels for the low eigenstates
are also approximately equally spaced with the energy difference
between the adjacent levels given by $\sqrt{NUJ}$.
\begin{figure}[h]
\includegraphics{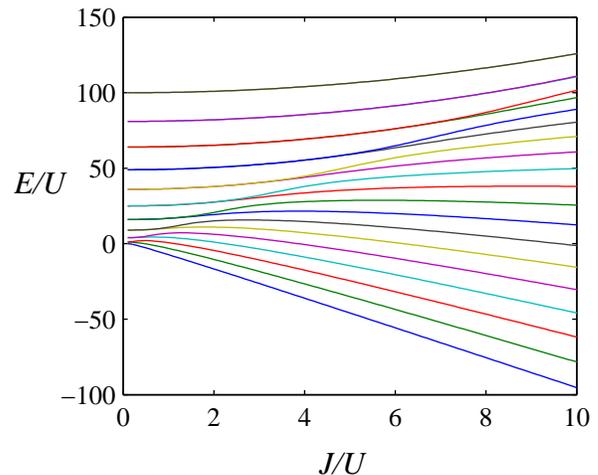}
\caption[Fig.2 ]{(Color Online) Energy spectrum of the Hamiltonian
(5) with the atom number N=20 \protect\cite{24}. }
\end{figure}

From the discussions above, we see that the Fock region has the
smallest energy spacing for the low-lying eigenstates with typical
experimental parameters. The non-adiabatic transition therefore
happens dominantly in or close to that region. However, due to the
quadratic scaling of the energy spacing in that region, although
non-adiabatic transition occurs, the atoms can only be excited to
some low-lying eigenstates if they start from the ground state.
The typical merging time in the experiment \cite{6} is about $0.5$
s, significantly longer than the time scale of $\hbar /\mu _{c}$,
where $\mu _{c}$ ($\sim 2$kHz) is the chemical potential. The
parameters $NU_{0}$ and $\mu _{c}$ are roughly on the same order
of magnitude for typical atomic gas
experiments. In the Fock region, the energy spacing between the $2\sqrt{N}$%
th excited state and the ground state is $NU_{0}$. So if the atoms
start from the ground state, the final spread in the energy
eigenstates due to the non-adiabatic transition is at most
$2\sqrt{N}$. Of course, in real experiments it is not practical to
start exactly from the ground state $\left| N/2,N/2\right\rangle $
of the Hamiltonian (5) when the traps are far apart. Let us assume
that the initial atom numbers $N_{1}$ and $N_{2}$ in the two traps
are roughly the same with $N_{1}\simeq N_{2}\simeq N/2$ \cite{22},
and that each of them has some fluctuations typically on the order
of $\sqrt{N}$ \cite {25}. Then, the initial spread in the energy
eigenlevel is about $2\sqrt{N}$. Adding together quadratically the
contributions from the non-adiabatic transfer and from the initial
number fluctuations , we conclude that after the merger, the atoms
will populate a mixture of the low-lying eigenstates of the
Hamiltonian (5), with a spread in the eigenstates characterized by
$2\sqrt{2N}$ (in terms of the order of magnitude).

Now we want to show that even if the final state after the merger
is a mixture of low-lying eigenstates, as long as the spread in
the eigenstates $2\sqrt{2N}\ll N$, we still have a large
single-condensate fraction. When the two traps are merged and we
are in the Rabi region, we can calculate the condensate fraction
in the mode $a_{c}=\left( a_{1}+a_{2}\right) /\sqrt{2}$
for each eigenstate of the $\hat{S_{x}}$ operator. For the $j$th eigenstate $%
\left| S=N/2,S_{x}=N/2-j\right\rangle $ of the $\hat{S_{x}}$
operator, its population in the mode $a_{c}$ is given by $N-j$
(This result can be easily seen if we rotate $\hat{S_{x}}$ to
$\hat{S_{z}}$ through a linear transformation on the modes $a_{1}$
and $a_{2})$. Now we are in a mixture of the eigenstates $\left|
S=N/2,S_{x}=N/2-j\right\rangle $ of the $\hat{S_{x}}$ operator
with $j$ up to $2\sqrt{2N}$. For this mixture state, the atom
number in the particular mode $a_{c}$ is at least $N-2\sqrt{2N}$.
Therefore, if the atom number $N$ is huge, we end up with the
dominant fraction of atoms in the mode $a_{c}$ even if the merging
is far outside the perfect adiabatic limit. The basic reason for
this result lies in the fast growth of the excitation energy with
the levels. The system is kept in some low-lying states of the
Hamiltonian (5) although the evolution is non-adiabatic. When most
of the atoms are in the condensate mode $a_{c}$, a unique relative
phase of zero is correspondingly established between the two
initial modes $a_{1}$ and $a_{2}$.

\section{Numerical Simulation of the merging dynamics of two condensates}

\subsection{Numerical simulation methods}

In the previous sections, we have shown that through a direct
merger it is possible to establish a unique relative phase between
two initially independent condensates, and argued that this
relative phase is robust and we can almost get a single condensate
even if the merging is far outside the adiabatic limit
($t_{m}\ll\hbar /U_{0}$). In this section, we would like to
quantitatively test this result through more detailed numerical
simulations of the merging dynamics.

To quantitatively describe the merging dynamics, we look at the evolution of
the largest condensate fraction derived from the single-particle density
matrix (6). The largest eigenvalue of the single-particle density matrix $%
\rho $\ can always be expressed as an expectation value $\left\langle
a_{c}^{\dagger }a_{c}\right\rangle $, where $a_{c}$ is a rotated mode
corresponding to the largest condensate fraction and is defined as $%
a_{c}=\cos \theta a_{1}+\sin \theta e^{i\varphi }a_{2}$. The rotation angle $%
\theta $ reflects the contributions to the new condensate mode $a_{c}$ from
the two initial modes $a_{1}$ and $a_{2}$, and $\varphi $ specifies their
relative phase in this new mode. With this characterization of the relative
phase, here we ignore explicit discussion on the inherent quantum
uncertainty of the phase operator caused by the finite total atom number.
This quantum uncertainty is on the order of $1/\sqrt{N}$, where the atom
number $N\sim 10^{6}$ for typical experiments, so it represents a small
effect. We refer the readers to Ref. \cite{phase} for detailed discussions
on that issue. Numerically, we start with certain initial states of the
modes $a_{1}$ and $a_{2}$, as will be detailed in the following subsections,
solve the evolution of these states under the Hamiltonian (5), and calculate
the single-particle density matrix for each instantaneous state to find out
the evolution of the atom number $\left\langle a_{c}^{\dagger
}a_{c}\right\rangle $ in the largest condensate fraction mode $a_{c}$ and
the corresponding parameters $\theta $ and $\varphi $. The atom number $%
\left\langle a_{c}^{\dagger }a_{c}\right\rangle $ is given by the
largest eigenvalue of $\rho $, while $\theta $ and $\varphi $ are
specified by the corresponding eigenvector $(\cos \theta ,\sin
\theta e^{i\varphi })$.

To quantify the state evolution, we need to specify the parameters $U\left(
t\right) $ and $J\left( t\right) $ in the Hamiltonian (5) as a function of
time. To be consistent with the experimental configurations in Ref. (13), we
assume that the two initial condensates are confined in cigar-shaped traps
which are moving towards each other along the radial direction with uniform
speed. As we have mentioned before, the details of the functions $U\left(
t\right) $ and $J\left( t\right) $ are not important, but rather their time
scales. So, for simplicity, we assume that the wave functions $\phi _{i}(%
\mathbf{r,t})$ $\left( i=1,2\right) $ of the modes $a_{i}$, which
adiabatically follow the movement of the cigar-shaped traps, also have
cigar-shaped profiles specified by the following Gaussian function

\begin{eqnarray}
\phi _{i}(\mathbf{r,t})=\frac{1}{\pi ^{3/4}\sigma _{\rho }\sqrt{\sigma }_{z}}
\exp \left[ -\frac{\left[ x-x_{i0}\left( t\right) \right] ^{2}+y^{2}}{
2\sigma _{\rho }^{2}}\right]  \nonumber \\
\times \exp \left[ -\frac{z^{2}}{2\sigma _{z}^{2}} \right] , \hspace*{3.5cm}
\label{12}
\end{eqnarray}
where $\sigma _{\rho }$ and $\sigma _{z}$ are the widths of the wave
functions along the radial and the axial directions, respectively. The
centers of the wavepackets are given by $x_{10}(t)=x_{0}\left(
1-t/t_{m}\right) /2$ and $x_{20}(t)=-x_{0}\left( 1-t/t_{m}\right) /2$,
defining a constantly diminishing separation from $x_0$ to $0$ throughout
the merging time $t_{m}$. The variation of the parameters $U\left( t\right) $
and $J\left( t\right) $ are then calculated with Eqs. (3)(4) from these
postulated wave functions. The typical evolution of $U\left( t\right)$ and $%
J\left( t\right)$ are shown in Fig. 5(a-b) with $\sigma _{z }=10\sigma
_{\rho}$, following the experimental conditions \cite{6}. Although in real
experiments the total atom number is normally on the order of $\sim 10^{6} $%
, in numerical simulations, due to the limited computation speed, we can
only deal with moderate atom numbers with $N\sim 10^{2}$. However, we know
that $NU_{0}$ and $J_{0}$ are typically on the same order of magnitude with $%
J_0$ somewhat smaller than $NU_0$. In order to be consistent with the
practical configurations, we should re-scale the parameters $U\left(
t\right) $ and $J\left( t\right) $ with a global constant so that $%
NU_{0}\sim J_{0}$. In our simulation, without loss of generality, we assume $%
NU_{0}=4 J_{0}$. When $U\left( t\right) $ and $J\left( t\right) $ are
specified, we can numerically calculate the evolution of the atom number $%
\left\langle a_{c}^{\dagger }a_{c}\right\rangle $ in the largest condensate
fraction and the relative phase $\varphi $ with the method described above.

\subsection{Merging of two condensates in Fock states}

We first investigate the merging of two condensates in Fock
states, with the atom numbers $N_{1}=51$ and $N_{2}=49$,
respectively. The numerical results are shown in Figs. (3a)-(3h),
where the largest condensate fraction $\eta $
(defined as $\eta =\left\langle a_{c}^{\dagger }a_{c}\right\rangle /N$ with $%
N=N_{1}+N_{2}$) and the parameters ($\theta $,$\varphi $) are
plotted against the dimensionless time $U_{0}t$. Note that for the
initial Fock state, the relative phase $\varphi $ is not well
defined (random). In the numerical program, one basically randomly
picks up a particular initial phase. In Fig. 3, the initial value
of $\varphi $ is set to zero simply due to the convention of the
program (it sets phase $\varphi =0$ for the complex number
$0e^{i\varphi }$). The merging dynamics and the final relative
phase do not depend on this particular choice. We test other
assignments of the initial phase, and basically there are no
changes in the dynamics (except for different initial jumps of the
phase which are meaningless). We have different
merging time scales for figures (a)-(h), ranging from the adiabatic limit $%
U_{0}t_{m}=40$ to the complete non-adiabatic limit $U_{0}t_{m}=0.04$, where $%
t_{m}$ denotes the time of the merger. With $U_{0}t_{m}=40\gg 1$,
the picture of the adiabatic mapping is basically valid, and as
expected, finally we have more than $95\%$ of atoms evolving into
the largest condensate fraction
(see Fig. 3g). For Figs. (3e) and (3c), the time scales change from $%
U_{0}t_{m}=4$ to $U_{0}t_{m}=0.4$, which is certainly outside the
adiabatic limit as $U_{0}t_{m}\lesssim 1$. However, since $%
NU_{0}t_{m}=100U_{0}t_{m}\gg 1$, as we analyzed in Sec. IIC, only
some low-lying eigenlevels will be populated during the evolution,
and we should have a good condensate fraction. This is confirmed
by the figures (3e) and (3c). Compared with Fig. (3g), the final
condensate fraction is reduced a little bit, but not much, in
particular for Fig. (3e) with $NU_{0}t_{m}\gg 1$. Finally, in Fig.
(3a) the variation is so fast ($NU_{0}t_{m}=4$) that many of the
eigenlevels of the Hamiltonian (5) will be populated and we expect
to have a poor final condensate fraction. This is confirmed by the
numerical simulation, where we can hardly see any considerable
increase of the condensate fraction.

Figures. (3b) (3d) (3f) and (3h) show the evolution of the
corresponding condensate mode which has the largest fraction of
the atoms. For Figs. (3d) (3f) and (3h), we expect that the final
state should be close to the lowest eigenstate of the operator
$-\hat{S}_{x}$. The corresponding condensate mode should be
$a_{c}\simeq \left( a_{1}+a_{2}\right) /\sqrt{2}$, with $\theta
\rightarrow \pi /4$ and $\varphi \rightarrow 0$. Indeed, we can
see from these three figures that the parameters $\theta $ and
$\varphi $ approach their stationary values with very small
oscillations as soon as $J\left( t\right) $ becomes comparable to
$U\left( t\right) $. When $J\left( t\right) \sim U(t)$, the system
is well inside the Josephson region. This means that a unique
relative phase has been established between the two initial modes
$a_{1}$ and $a_{2}$ when their coupling is still weak. In such a
region, the two-mode model is still well justified, which supports
the application of the two-mode model in understanding the
establishment of the relative phase. The signature from Fig. (3b)
is less clear because the corresponding increase in the condensate
fraction still remains to be small.
\begin{figure*}[tbp]
\includegraphics{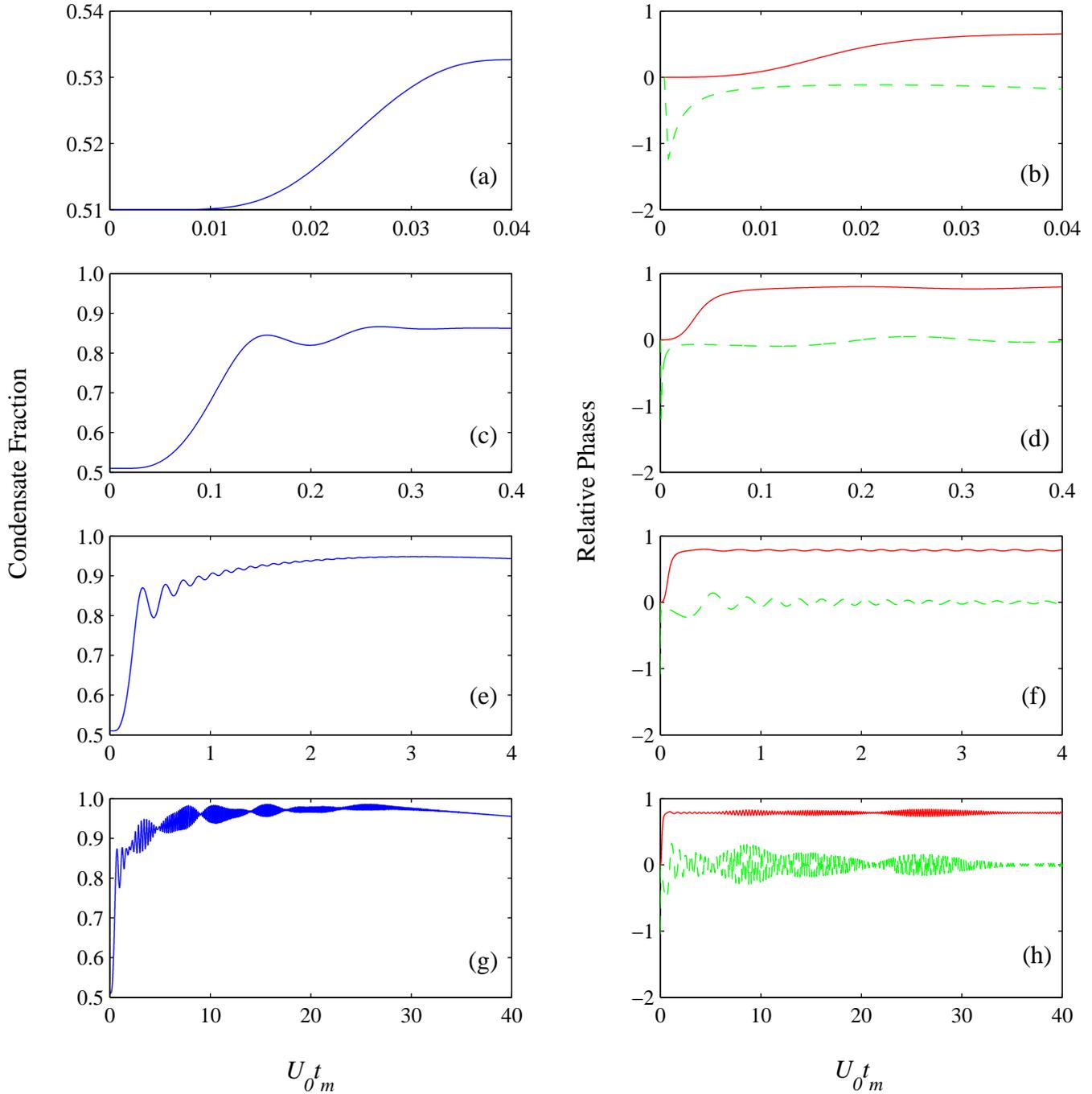}
\caption[Fig.3 ]{(Color Online) Evolution of the largest
condensate fraction and the relative phase parameters for various
merging time scales represented by the dimensionless parameter
$U_{0}t_{m}$. Figs. \textbf{(a),(c),(e),(g)} are for
evolution of the largest condensate fractions, with the merging time scales $%
U_{0}t_{m}=0.04$, $U_{0}t_{m}=0.4$, $U_{0}t_{m}=4$, $U_{0}t_{m}=40$,
respectively. Figures \textbf{(b),(d),(f),(h)} are for evolution of the
rotation angle $\protect\theta $ (solid line) and the relative phase $%
\protect\varphi $ (dashed line), with the merging time scales corresponding
to Figs. \textbf{(a),(c),(e),(g)}, respectively.}
\end{figure*}

\subsection{Merging of two condensates in Fock and coherent states}

In this subsection, we investigate the merging of two condensates in a Fock
state with $N_{1}=50$ and in a coherent state $\left| \alpha \right\rangle $
with $\left| \alpha \right| ^{2}=64$, respectively. This is motivated, on
the one hand, by the consideration that the source for an operating atom
laser resembles a coherent state; and on the other hand, by the curiosity to
find out the influence of the initial number fluctuations on the merging
dynamics. A coherent state is certainly not an eigenstate of the Hamiltonian
(5) at $t=0$ with $J\left( t\right) \rightarrow 0$, so we start from a
superposition of a series of eigenlevels instead of a particular one.

The simulation results are shown in Figs. (4a)-(4e) for the evolutions of
the largest condensate fraction $\eta $ and the corresponding parameters ($%
\theta $, $\varphi $), with the merging time varying from $U_0 t_m=0.04$ to $%
U_0 t_m=4$. The results are qualitatively similar to the ones displayed in
Fig. 3. The main difference is that the final condensate fraction
corresponding to the same time scale is notably worse now. This is
understandable, as the initial state we start with is not an eigenstate of
the governing Hamiltonian and it has significant initial number fluctuation.
The coherent state has number fluctuation on the order of $\sqrt{N}$, so
from the analysis in Sec. IIC, we expect that the final condensate fraction
may reduce by an amount of order $\sqrt{N} /N$, which is pretty close to the
results shown in Figs. (4a)-(4e). In our simulation, the coherent state is
not large due to the limitation of the computation efficiency; $1/\sqrt{N}$
is therefore not a small factor and we see considerable decrease in the
condensate fraction. In more practical cases with $N\sim 10^{6}$, the
initial number fluctuation of each condensate should have a much smaller
influence on the final condensate fraction.
\begin{figure*}[tbp]
\includegraphics{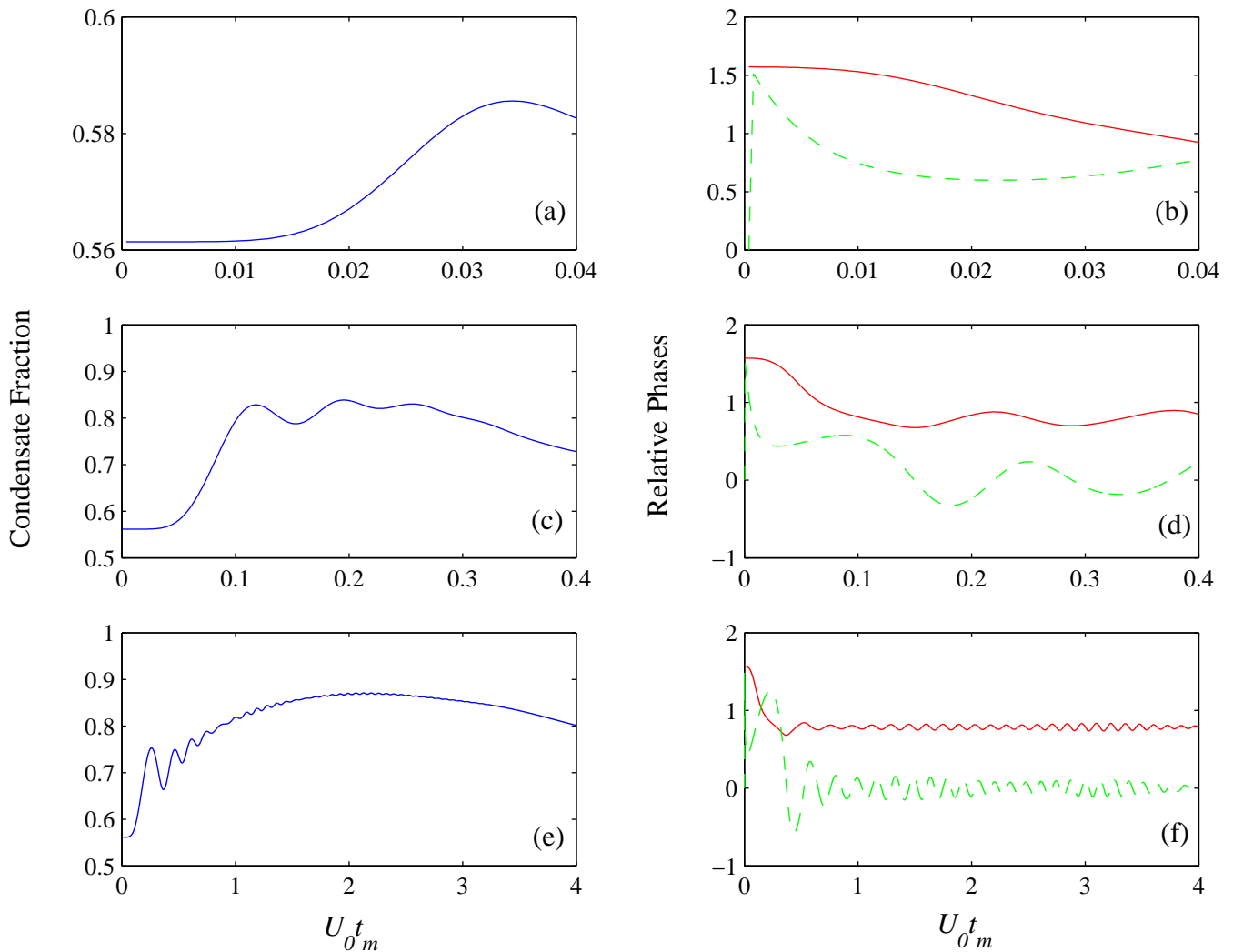} .
\caption[Fig.4 ]{(Color Online) Similar to Fig.3, but with
different states for the initial component condensates. One of
them is in a Fock state with N=50 and the
other is in a coherent state $\left| \protect\alpha \right \rangle$ with $%
\protect\alpha =8$. The merging time scales are given respectively by $%
U_{0}t_m=0.04$ (Figs. (a),(b)), $U_{0}t_m=0.4$ (Figs. (c),(d)), $U_{0}t_m=4$
(Figs. (e),(f))}
\end{figure*}

\subsection{Comparison of different merging methods}

We have mentioned before that the merging dynamics is only determined by the
rough time scale in the variations of $U\left( t\right) $ and $J\left(
t\right) $, rather than the explicit functional forms of these two
parameters. We now verify this result by considering different merging
methods. Let us consider two cigar-shaped condensates (with $\sigma _{z
}=10\sigma _{\rho}$ in Eq. (7)) merged along the radial and the axial
directions, respectively. The variations of $U\left( t\right) $ and $J\left(
t\right) $ as functions of time are shown in Fig. (5a) and (5b) for these
two cases. For merging along the axial direction, the evolutions of the
largest condensate fraction $\eta $ and the corresponding mode parameters ($%
\theta $,$\varphi $) are shown in Fig. (5c) and (5d) with $U_{0}t_{m}=4$.
Compared with the corresponding results for merging along the radial
direction, we see there is little difference in the final condensate
fraction although the variations of $U\left( t\right) $ and $J\left(
t\right) $ differ considerably in the two cases. This shows that what
matters most for the merging dynamics is the rough time scale of $U\left(
t\right) $ and $J\left( t\right) $ as we have mentioned before. It also
justifies our use of the simple Gaussian functions in Eq. (7) to model the
individual condensate wave functions, as the functional details of $U\left(
t\right) $ and $J\left( t\right) $ are not so important.
\begin{figure*}[tbp]
\includegraphics{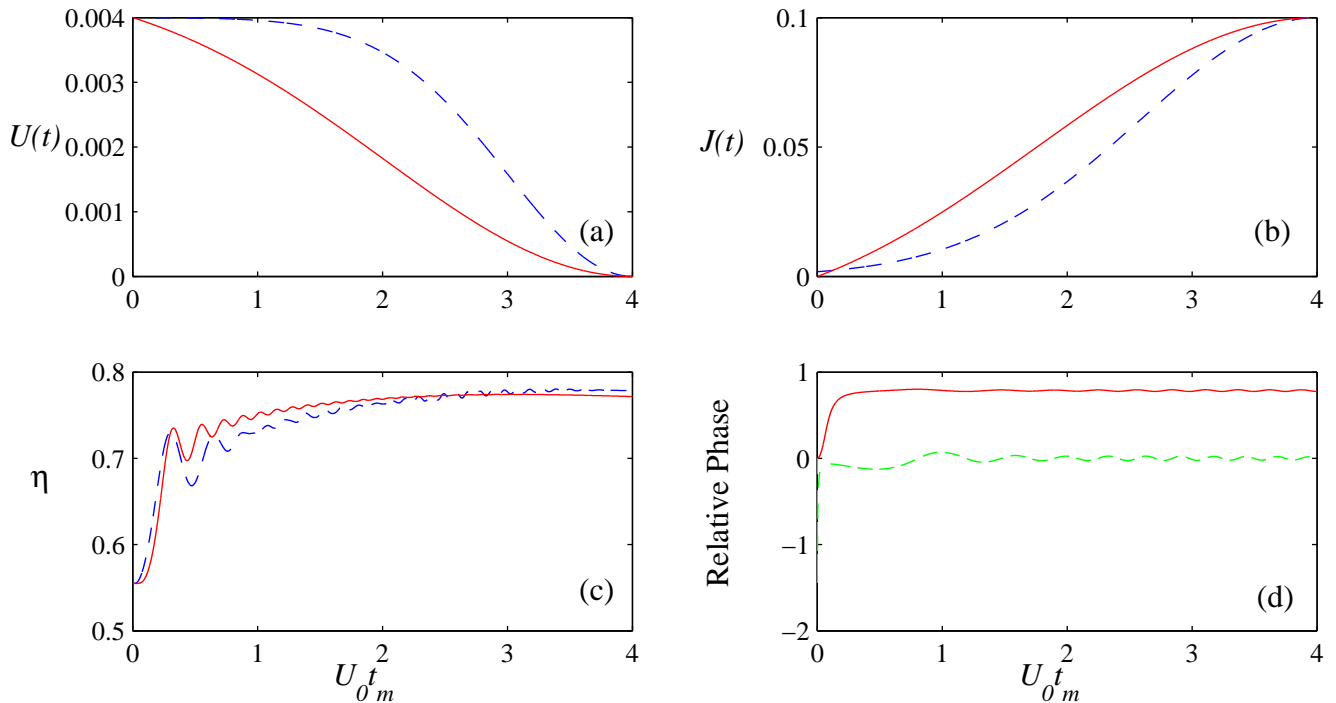}
\caption[Fig.5 ]{(Color Online) Comparison of the dynamics for
merging along different directions. \textbf{(a)(b)} $U$ and $J$ as
functions of the dimensionless time while merging along the radial
direction (solid line) or the axial direction (dashed
line).\textbf{(c)} Comparison of the evolution of the largest
condensate fraction $\protect\eta$. The solid line represents
merging along the radial direction and the dashed line represents
merging along the axial direction. Despite different variations of
$U$ and $J$ shown in (a)(b), there seems to be little difference
in the evolution of the
condensate fraction. \textbf{(d)} Evolution of the rotation angle $\protect%
\theta $(solid line) and the relative phase $\protect\varphi $(dashed line)
when merging along the the axial direction. Qualitatively, it is similar to
the evolution shown in Fig. 3(f) for merging along the radial direction.}
\end{figure*}

This result should not be misunderstood as that it is equally good
in real experiments to merge the condensates along the radial or
the axial direction. Experimentally, for the cigar-shaped
condensates, it is much better to merge them along the radial
direction \cite{6}. The reason is that we have assumed that the
density profile of the atoms can adiabatically follow the
movements of the traps to validate the two-mode model. Although
that is the case for merging along the radial direction, it will
be much harder to meet this condition if the condensates are
merged along the axial direction. In the latter case, one needs to
have the condensates further away from each other initially to
have negligible $J\left( t\right) $ at the beginning; one also
needs to move the atoms significantly faster to have the same time
scale for merging. However, the trap along the axial direction is
much weaker, and a large fraction of the atoms could be excited
during the merger. As a result, the atomic density profile would
be left behind the movements of the traps unless one reduces the
merging speed to an undesirable value (for instance, with a time
longer than the condensate life time). Therefore, merging along
the weaker trapping direction makes the individual condensate wave
functions harder to follow adiabatically the movements of the
traps, which would invalidate the two-mode approximation.
Nevertheless, if the two-mode approximation could be justified,
the relative phase dynamics would then become insensitive to the
detailed merging methods, as shown by this numerical simulation.

\section{Summary}

Using a two-mode model, we have studied the dynamics of relative phase and
condensate fraction during the direct merging of two independently prepared
condensates with random relative phase. In accordance with a recent
experiment\cite{6}, we found that it is possible to create a single
condensate with larger condensate fraction and a unique zero relative phase
between the initial condensate modes. By examining the energy spectrum of
the Hamiltonian under the two-mode approximation, the process can be
understood both within and without the adiabatic limit. In the ideal
adiabatic limit, the result can be explained using the adiabatic theorem,
and the system will evolve from a fragmented condensate to a single
condensate following the evolution of the eigenstate of the Hamiltonian
during the merger. Beyond the adiabatic limit, careful analysis of the
evolution of the eigen-spectrum is needed. Qualitatively, due to the
quadratic scaling of the excitation energy with the energy levels, only the
low-lying eigenstates of the Hamiltonian will be populated even if the
process is far from the adiabatic limit. At the end of the merger, the
mixture of those low-lying states will give rise to the final state with
dominant number of atoms in the desired single condensate mode. Numerical
simulations are then performed, and the results are in good agreement with
our analysis. The results may have interesting implications for realization
of a continuous atom laser based on direct merging of independently prepared
condensates. Because of the similarity of our model to the case of
condensates in double wells or optical lattice, the adiabatic and
non-adiabatic evolution picture described here may also find applications in
controlling the dynamics of condensates in those optical potentials.

This work was supported by the ARDA under ARO contracts, the NSF EMT grant,
the FOCUS seed funding, and the A. P. Sloan Fellowship.


\begin{thebibliography}{99}
\bibitem{1}  M. Kozuma \textit{et al.}, \textit{Science}, \textbf{286}, 2309
(1999).

\bibitem{2}  S. Inouye \textit{et al.}, \textit{Nature}, \textbf{402}, 641
(1999).

\bibitem{16}  M.-O. Mewes \textit{et al.}, \textit{Phys. Rev. Lett.},
\textbf{78}, 582 (1997).

\bibitem{3}  C.K. Law, N.P. Bigelow, \textit{Phys. Rev. A}, \textbf{58},
4791-4795 (1998).

\bibitem{13}  M. Holland, K. Burnett, C. Gardiner, J.I. Cirac and P. Zoller,
\textit{Phys. Rev. A}, \textbf{54}, R1757 (1996).

\bibitem{14}  R.J.C. Spreeuw, T. Pfau, U. Janicke, and M. Wilkens, \textit{%
Eur. Phys. Lett.}, \textbf{32}, 469 (1995).

\bibitem{15}  H.M. Wiseman, M.J. Collett, \textit{Phys. Lett. A}, \textbf{202%
}, 246 (1995).

\bibitem{17}  J. Williams, R. Walser, C. Wieman, J. Cooper, and M. Holland,
\textit{Phys. Rev. A}, \textbf{57}, 2030 (1998).

\bibitem{18}  B.K. Teo, G. Raithel, \textit{Phys. Rev. A},\textbf{63},
031402 (2001).

\bibitem{19}  M. Greiner \textit{et al.}, \textit{Phys. Rev. A}, \textbf{63}%
, 031401 (2001).

\bibitem{4}  D. Jaksch \textit{et al.}, \textit{arXiv:cond-mat/0101057}
(2001).

\bibitem{5}  K. M{\o }lmer, \textit{Phys. Rev. A}, \textbf{65}, 021607
(2002).

\bibitem{6}  A.P.Chikkatur \textit{et al.}, \textit{Science}, \textbf{296},
2193 (2002).

\bibitem{7}  I. Zapata, F. Sols, and A.J. Leggett, \textit{Phys. Rev. A},
\textbf{67}, 021603 (2003).

\bibitem{8}  J. Javanainen, M.Y. Ivanov, \textit{Phys. Rev. A}, \textbf{60},
2351 (1999).

\bibitem{9}  K.W. Mahmud, H. Perry, and W.P. Reinhardt, \textit{%
arXiv:cond-mat/0312016} (2003).

\bibitem{10}  R.W. Spekkens, J.E. Sipe, \textit{Phys. Rev. A}, \textbf{59},
3868 (1999).

\bibitem{11}  G.J. Milburn, J. Corney, E.M. Wright, D.F. Walls, \textit{%
Phys. Rev. A}, \textbf{55}, 4318, (1997).

\bibitem{12}  C. Menotti, J.R. Anglin, J.I. Cirac and P. Zoller, \textit{%
arXiv:cond-mat/0005302} (2000).

\bibitem{21}  A.J. Leggett, \textit{Rev. Mod. Phys.}, \textbf{73}, 307
(2001).

\bibitem{20}  D. Wagner, \textit{J. Phys. A: Math. Gen.}, \textbf{15}, 3307
(1982).

\bibitem{22}  We have assumed that the two traps are initially symmetric. In
this case, the system is in a low-energy state only when $N_{1}\simeq N_{2}$%
. It is possible to generalize the results to the case where the initial two
traps are asymmetric and $N_{1}\neq N_{2}$ for the low energy state.

\bibitem{23}  V.K. Thankappan, \textit{Quantum Mechanics} (Pergamon Press,
New York 1993).

\bibitem{25}  It is very hard to control the initial atom number of each
condensate to be exactly the same ($N/2$). We assume there is a
number flucation on the order of $\sqrt{N}$ around this mean
value, corresponding to the so-called shot-noise limit.

\bibitem{24}  M. Holthaus, S. Stenholm, \textit{Eur. Phys. J. B}, \textbf{20}%
, 451 (2000).

\bibitem{phase}  G-S. Paraoanu \textit{et al.}, \textit{J. Phys. B: At. Mol.
Phys.}, \textbf{34}, 4689 (2001).
\end{thebibliography}
\end{document}